# Effect of coarse-graining on detrended fluctuation analysis

Radhakrishnan Nagarajan*

*Center on Aging, University of Arkansas for Medical Sciences*

**Abstract**

Several studies have investigated the scaling behavior in naturally occurring biological and physical processes using techniques such as Detrended Fluctuation Analysis (DFA) [1]. Data acquisition is an inherent part of these studies and maps the continuous natural process into a digital data. The resulting digital data is discretized in amplitude and time, and shall be referred to as *coarse-grained realization* in the present study. Since coarse-graining precedes scaling exponent analysis, it is important to understand its effects on scaling exponent estimators such as DFA. In this brief communication, k-means clustering is used to generate coarse-grained realizations of data sets with different correlation properties, namely: anti-correlated noise, long-range correlated noise and uncorrelated noise. It is shown that the coarse-graining can significantly affect the scaling exponent estimates. It is also shown that scaling exponent can be reliably estimated even at low levels of coarse-graining and the number the clusters required varies across the data sets with different correlation properties.

**Keywords**: detrended fluctuation analysis, coarse-graining, k-means

**PACS Code** : **05.40.-a**

*To whom correspondence should be addressed*

Radhakrishnan Nagarajan
Center on Aging, University of Arkansas for Medical Sciences
629 Jack Stephens Drive, Room: 3105
Little Rock, Arkansas 72205
Phone: (501) 526 7461
Email: nagarajanradhakrish@uams.edu

## 1. Introduction

Detrended fluctuation analysis (DFA) has been used to understand the nature of correlations in data recorded from a wide range of biological and physical systems. Acquiring data from natural processes involves several intermediate steps. A schematic diagram of the data acquisition procedure is shown in Figure 1 [2]. The transducer or the measurement device converts the natural process into a continuous-time data $x_t, t \in R, x_t \in R$. The amplitude of the continuous-time data is proportional to the magnitude of the process, hence representative of the underlying dynamics. The continuous-time data $x_t$ in turn is converted into a discrete-time data $x_n$, $t = nT = n/F_s$ by sampling at a specified sampling frequency $F_s$, represented by the second block in Figure 1. For narrow-band signals it is important to choose the sampling frequency so as to satisfy the Nyquist criterion [2]. However, such a constraint may not be necessary for power-law processes, as they are broadband in nature. The discrete time data in turn is converted into a discrete-valued data $x_n^q$ by a quantizer, represented by the third block in Figure 1. In the following sections, we shall refer to this discretized data $x_n^q$ as *coarse-grained* realization of the actual process. Coarse-graining can be regarded as sampling of the probability density function f(x) of $x_t, t \in R$. In the present study, k-means clustering algorithm [3, 4], is used to map the amplitude of the continuous probability density function f(x) as a sum of dirac delta distributions $\sum_{i=1}^{n_c} d(x - q_i)$, where $q_i$ represent the center of mass (centroids) of the elements in the $i^{th}$ cluster, $i = 1...n_c$ (Section 2). In essence, the given data is mapped into a sequence with alphabet set of size $n_c$. From the above expression it is clear that increasing the number of levels $n_c$ yields better representation of the distribution, hence finer *resolution* of the given data. However, the focus of the present study is to investigate the effect of the number of levels ($n_c$) on retaining the *dynamical characteristics* as opposed to the *signal strength* given by the signal to noise ratio (Section 2). This is attributed to the fact that a finer resolution may not necessarily be required to capture the dynamical characteristics of the given data [5]. Coarse-grained realizations or symbolic reconstructions of deterministic nonlinear dynamical systems have been found to mirror the transformation between adjacent states in the real space by shift operators in the symbol space [5]. However, such rigorous mathematical formulations cannot be readily extended to stochastic processes.

In the present study, we investigate the effect of coarse-graining on the dynamical characteristics, hence the *scaling behavior* of the given data. We show that reliable estimate of the scaling exponents can be obtained even at lower resolutions as reflected by ($n_c$), and the required resolution varies across anti-correlated, uncorrelated and long-range correlated data. This report is organized as follows: in Section 2, k-means clustering is used to generate coarse-grained realization of the given data. In Section 3, the effect of coarse-graining on scaling exponent estimates of anti-correlated noise, uncorrelated noise and long-range correlated noise is discussed. Section 3 is followed by a discussion, Section 4. The data sets considered in this study are publicly available [6].

**2. Coarse-graining using k-means clustering technique**

In the present study, we use the k-means clustering technique [3, 4] to generate coarse-grained realizations of the given data. Each sample in the given data must belong to a cluster and belongs to only one cluster, hence k-means partitions the given data into a specified number of clusters, $n_c$. The k-means algorithm does not assume any specific distribution of the given data and hence highly suitable for generating coarse-grained realizations of experimental data sets. The algorithm is iterative and groups the given data into a specified number of $n_c$ clusters using a suitable metric. In the present study we use the $l_2$-norm or Euclidean metric. It is well known that the k-means technique is sensitive to the initial choice of centroids. There have been several articles on the initial choice of the centroids [9]. The ad-hoc approach is to repeat the clustering procedure with several different initializations. To alleviate these issues, preliminary clustering was performed using 10% of the samples with random initialization. The resulting centroids in turn were used as initial guess for clustering the entire data. The algorithm converges when there is no appreciable movement of the centroids [4]. The coarse-grained data is given by $x_i^q = \mathbf{m}(c_k)$ if $x_i \in c_k$, $k = 1...n_c$, $i = 1...n_c$, where $\mathbf{m}(c_k)$ represents the centroid of cluster $c_k$. The error as a result of coarse-graining is given by $x_i^e = x_i - x_i^q, i = 1...n$. The *signal to noise ratio* is given by the expression SNR (dB) = $20\log_{10}(\frac{\mathbf{s}_x}{\mathbf{s}_e})$ where $\mathbf{s}_x$ and $\mathbf{s}_e$ represent the standard deviation of the given data $x$ and the noise $x^e$ respectively. From the above expression, it is evident that smaller variance of the noise results in larger SNR, hence better representations of the given data. The SNRs of the anti-correlated (α = 0.2), uncorrelated (α = 0.5) and long-range correlated noise (α = 0.8) with $n_c = $ 2, 4, 6, 8, 10, 12, 14, 16, 18 and 20 clusters are shown in Figure 2. The plot of SNR versus $\log_2(n_c)$ exhibited a linear trend ($y = m\,x + c$) whose parameters (*m, c*) were similar across three data sets, Figure 2. Thus the SNR is immune to the nature of correlation in the given data and is directly proportional to $\log_2(n_c)$. This is also evident from the expression for SNR which is governed solely by the probability distribution (variance) of the given data. Symbolic reconstruction $x^s$ of the given data can be achieved by assigning a symbol $s_i, i = 1...n_c$ to each of the cluster centroids $c_i, i = 1...n_c$ where the clusters represent the alphabet set of the resulting sequence. For linear processes, the spectral power, hence the correlation properties of $x^s$ is determined from the corresponding *indicator sequences* [10]. The indicator sequence $I_{s_i}$ corresponding to symbol $s_i$ is generated as

$$I_{s_i}(k) = 1, \text{ if } I_{s_i}(k) = s_i, k = 1...n, i = 1...n_c$$
$$= 0, \text{ otherwise}$$

Each of the indicator sequence is orthogonal to the other and the spectral power of the sequence $x^s$ is obtained by superposition of the spectral power across the $n_c$ indicator sequences. Such an approach has been useful in identifying tandem repeats and scaling behavior ($\beta = 2\alpha - 1$) in DNA fragments [11]. In a related note, DNA sequences can be thought of as coarse-grained versions of real data mapped into a symbolic sequence with alphabet size four represented by the nucleotides (A, G, C and T).

**3. Results**

In the present study, detrended fluctuation analysis with first order polynomial detrending (DFA-1) [1] is used to estimate the scaling exponent of the original and coarse-grained versions of anti-correlated noise ($\alpha = 0.2$), long-range correlated noise ($\alpha = 0.8$) and uncorrelated noise ($\alpha = 0.5$). The length was fixed at ($N = 2^{17}$) for all the data sets [6].

*(A) Anti-correlated noise* ($\alpha = 0.2$): For the anti-correlated noise ($\alpha = 0.2$), the coarse-grained data $x^q$ were generated with ($n_c$ = 2, 4, 6, 8, 10, 12, 14, 16, 18 and 20) clusters. The log-log plot ($\log_{10}$) of the fluctuation functions versus time scale of $x^q$ is shown in Figure 3. The slope of the log-log plot exhibited a decreasing trend towards that of the original data with increasing number of clusters. The maximum slope estimated by linear regression was observed for ($n_c$ = 2, $\alpha \sim 0.47$) resembling that of uncorrelated noise. For cluster size ($n_c$ = 20) the slope of $x^q$ resembles that of the original data ($\alpha \sim 0.2$) with maximum discrepancy for ($n_c$ = 2), Figure 4a. The overlap in the fluctuation function versus time scale for ($n_c$ = 20) is mirrored by the overlap in the power-spectrum for ($n_c$ = 20), Figure 4b. Therefore, for anti-correlated noise at least ($n_c$ = 20) clusters ($\lfloor \log_2(20) \rfloor + 1 \sim 5$ bits per sample) which corresponds to SNR ~ 22 dB, Figure 2, is required for reliable estimation of the scaling exponent. The power spectrum of the noise exhibits a characteristic trend similar to that of the anti-correlated noise ($\alpha$ = 0.2) for ($n_c$ = 2), Figure 5a. However, it becomes relatively uniform for ($n_c$ = 20), Figure 5a, indicating that the noise is uncorrelated with the coarse-grained data at this resolution ($n_c$ = 20). There is also a considerable reduction in the magnitude of the power for the noise from ($n_c$ = 2) to ($n_c$ = 20) indicating that its contribution to the scaling exponent is negligible at this resolution.

*(B) Long-range correlated noise* ($\alpha$ = 0.8): A similar analysis was carried out for long-range correlated noise ($\alpha$ = 0.8). The log-log plot ($\log_{10}$) of the fluctuation functions versus time scale for the coarse-grained data sets $x^q$ obtained using ($n_c$ = 2, 4, 6, 8, 10, 12, 14, 16, 18 and 20) clusters is shown in Figure 6. As with the case of anti-correlated noise, the discrepancy in the scaling exponent estimate decreased with increasing resolution ($n_c$). However, the slope of the log-log plot exhibited an increasing trend towards the original data ($\alpha$ = 0.8) with increasing number of clusters, Figure 6 with minimum slope for ($n_c$ = 2). This behavior is in contrast to that of anti-correlated noise, Figure 3, whose slope was maximum at ($n_c$ = 2) and exhibited a decreasing trend towards the true slope with increasing resolution ($n_c$). The log-log plot of the fluctuation function versus time scale obtained for $x^q$ with ($n_c$ = 2 and $n_c$ = 8) is shown in Figure 7a. A significant overlap in the log-log plot was observed between the original and the coarse-grained data for ($n_c$ = 8), also reflected in the corresponding power spectrum for ($n_c$ = 8), Figure 7b. Therefore, for long-range correlated noise at least ($n_c$ = 8) clusters ($\log_2(8) = 3$ bits per sample) which corresponds to SQNR ~ 14 dB, Figure 2, is required for reliable estimation of the scaling exponent. The power spectrum of the noise exhibited a characteristic trend similar to that of ($\alpha = 0.8$) for ($n_c = 2$), Figure 5b. However, it becomes relatively uniform for ($n_c = 8$), Figure 5b, indicating that the noise is uncorrelated with the coarse-grained realization at this resolution. There is also a considerable reduction in the magnitude of the power for the noise from ($n_c = 2$) to ($n_c = 8$) indicating that its contribution to the scaling exponent is negligible at this resolution.

*(C) Uncorrelated noise (a = 0.5)*: A similar analysis on uncorrelated noise was carried out. The log-log plot ($\log_{10}$) of the fluctuation functions versus time scale for $x^q$ obtained using ($n_c = 2$, 4, 6, 8, 10, 12, 14, 16, 18 and 20) clusters is shown in Figure 8. A considerable overlap in the slopes was between the coarse-grained realization $x^q$ and the original data for ($n_c = 4, 6, 8, 10, 12, 14, 16, 18$ and $20$). The slope corresponding to ($n_c = 2$) was similar to that of the original data but was shifted by an offset. Therefore, for uncorrelated noise at least ($n_c = 2$) clusters ($\log_2(2) = 1$ bit per sample) which corresponds to SQNR ~ 4 dB, Figure 2, is required for reliable estimation of the scaling exponent.

Recent reports have emphasized the importance of understanding sign and magnitude correlations in the given data [12-14]. It is important to note that the sign and magnitude series are essentially nonlinear transforms of the given data hence can significantly alter its correlation properties. The sign series basically maps increments of the given data into a sequence with two symbols similar to a binary partition ($n_c = 2$) discussed earlier. However, unlike the clustering technique, which partitions the probability distribution of the given data, the sign series is a *slope detector* and represents the *envelope* of the given data.

*(D) Qualitative assessment of Sign and Magnitude Correlations with coarse-graining*

The *sign series* [12, 13] $y_{si}, i = 1...n-1$ of the given data $x_i, i = 1...n$ is given by

$$y_{si} = 1, \text{if } x_{i+1} > x_i$$
$$= -1, \text{otherwise}$$

The corresponding *magnitude series* or *volatility* [12, 13] $y_{vi}, i = 1...n-1$ is given by

$$y_{vi} = x_{i+1} - x_i, \text{if } x_{i+1} > x_i$$
$$= x_i - x_{i+1}, \text{otherwise}$$

For the anti-correlated noise ($\alpha = 0.2$), scaling of the sign series resembles that of the coarse-grained counterpart ($n_c = 2$), Figure 9a with a scaling exponent ($\alpha \sim 0.5$) in the asymptotic regime (time scale $10^{1.5}$ to $10^3$), distinct from the true scaling exponent ($\alpha = 0.2$). Scaling of the magnitude series, Figure 9b, obtained with ($n_c$ = 2, 4, 6, 8, 10, 12, 14, 16, 18 and 20) exhibited ($\alpha \sim 0.5$) characteristic of monofractal data [12-14] even for low number of clusters ($n_c = 2$). Thus scaling behavior of the magnitude series, Figure 9b, unlike the original data, Figure 3, is robust to coarse-graining. Similar analysis of the long-range correlated noise ($\alpha \sim 0.8$) revealed distinct scaling behavior for the sign series ($\alpha \sim 0.5$) and the coarse-grained counterpart ($n_c = 2$, $\alpha \sim 0.8$), Figure 10a. This has to be contrasted with that of the anti-correlated data where the sign series and the coarse-grained counterpart ($n_c = 2$) revealed similar scaling behavior in the asymptotic regime. The magnitude series of the coarse-grained realizations, Figure 10b, revealed exponent close to ($\alpha \sim 0.5$) characteristic of monofractal data [12-14].

**4. Discussion**

DFA and its extensions have been used to determine the nature of correlations in data sets obtained from a wide variety of nature processes. Data acquisition is an important step prior to scaling exponent estimation. Such data sets represent the quantized or coarse-grained versions of the actual process. In the present study we discussed the impact of coarse-graining on the correlation properties of the given data. Coarse-graining was implemented with k-means clustering where the number of clusters determines the resolution. The effect of coarse-graining was investigated on data sets with different correlation properties, namely: anti-correlated noise, long-range correlated noise and uncorrelated noise. While the signal to noise ratio for these data sets were similar with increasing number of clusters their dynamical characteristics were quite different. Scaling exponent estimation of uncorrelated noise ($\alpha = 0.5$, $N = 2^{17}$) was immune to coarse-graining and a binary partition ($n_c = 2$, 1 bit/sample) was sufficient to extract the exponent. Long-range correlated noise ($\alpha = 0.8$, $N = 2^{17}$) required a minimum of 3 bits/sample ($n_c = 8$) for successful estimation of their scaling exponent. Anti-correlated noise ($\alpha = 0.2$, $N = 2^{17}$), required a minimum of 5 bits/sample ($n_c = 20$) for successful estimation of the scaling exponent. For lower resolutions the scaling exponent of the anti-correlated noise resembled that of uncorrelated noise. Scaling of the sign series resembled that of its coarse-grained counterpart ($n_c = 2$) for anti-correlated noise with exponent ($\alpha \sim 0.5$). However, this was not true for long-range correlated noise. The magnitude series of the data sets with various levels of coarse-graining revealed scaling exponent ($\alpha \sim 0.5$) for the anti-correlated and long-range correlated noise characteristic of monofractal data. Thus scaling analysis of the magnitude series can be robust to coarse-graining. The present study clearly demonstrates that coarse-graining can significantly affect the scaling exponent estimates and the required resolution varies across data sets with different correlation properties.

## 5. Acknowledgements

We would like to thank the reviewers for their useful comments and suggestions. We would also like members of Physionet for making available their published data.

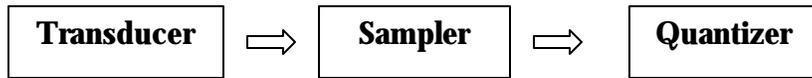

**Figure 1** Schematic diagram representing the elements of a typical data acquisition system. The transducer maps the physical process into a continuous time data $x_t$. The output of the transducer in turn is mapped into a discrete time signal $x_n$ by the sampler. The output of the sampler is in turn converted into a discrete-time discrete-valued data (coarse-grained data) $x_n^q$ by the quantizer.

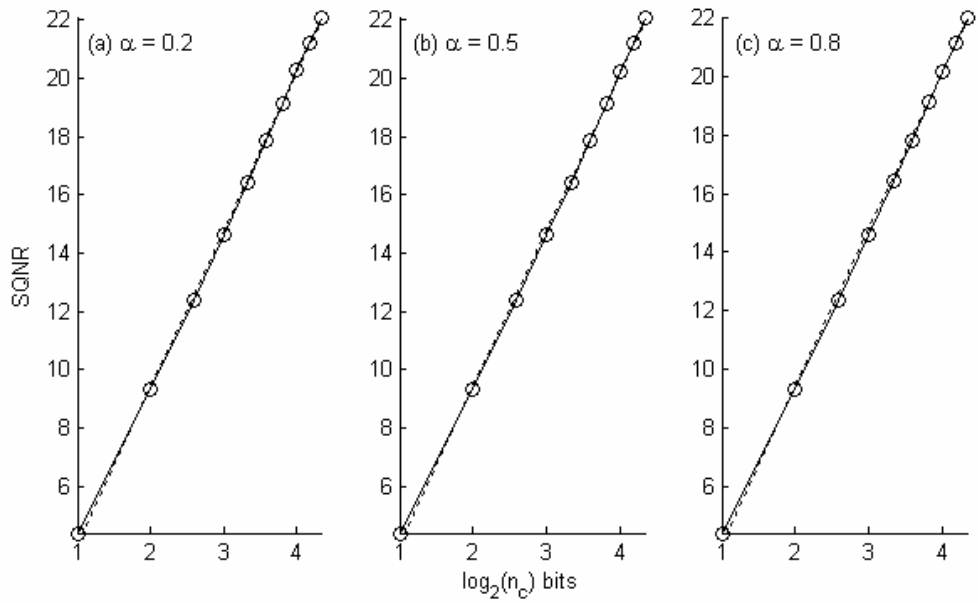

**Figure 2** Variation of the signal to noise ratio (SNR) with respect to the number of bits $\log_2 n_c$ where $n_c$ = 2,4,6,8,10,12,14,16,18 and 20, (solid lines) for the anti-correlated noise ($\alpha$ = 0.2), uncorrelated noise ($\alpha$ = 0.5), and long-range correlated noise ($\alpha$ = 0.8), shown in (a), (b) and (c) respectively. The slopes of the curves in (a), (b) and (c) are similar and of the form $y = m\,x + c$, where (m ~ 5.4, c ~ -1.3) (dotted lines).

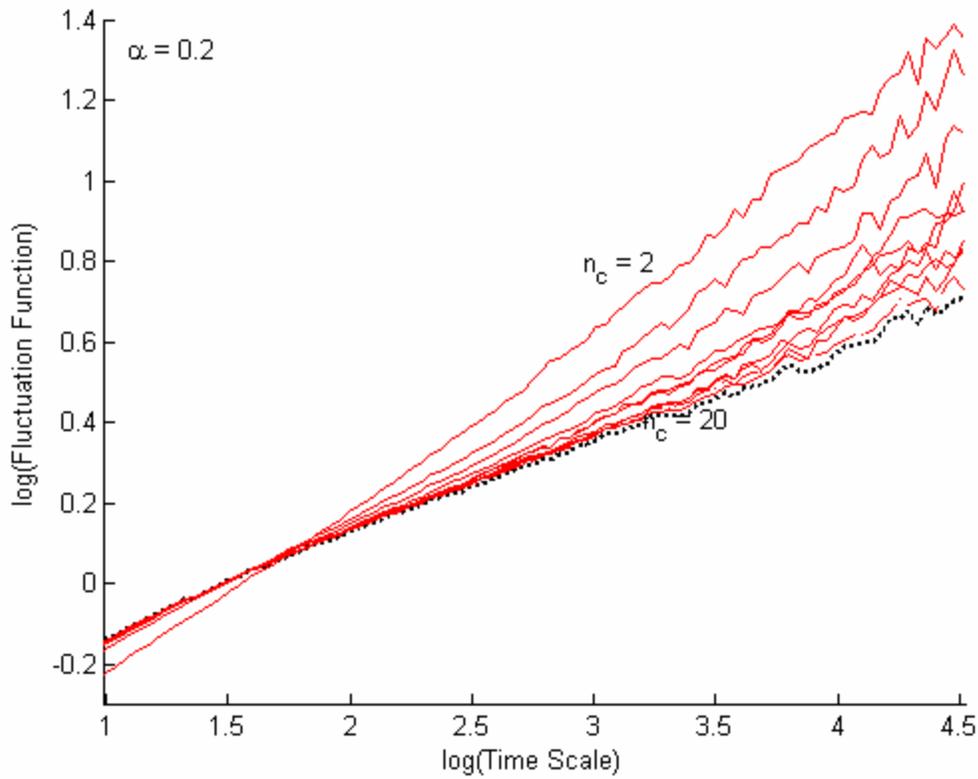

**Figure 3** Log-log plot of the fluctuation function versus time-scale of the anti-correlated noise ($\alpha = 0.2$, dotted line) and its coarse-grained versions (solid lines) generated for clusters ($n_c = 2, 4, 6, 8, 10, 12, 14, 16, 18$ and $20$). The scaling of the coarse-grained realizations exhibit an decreasing trend with increasing number of clusters and converge to that of the actual data ($\alpha = 0.2$) for ($n_c = 20$).

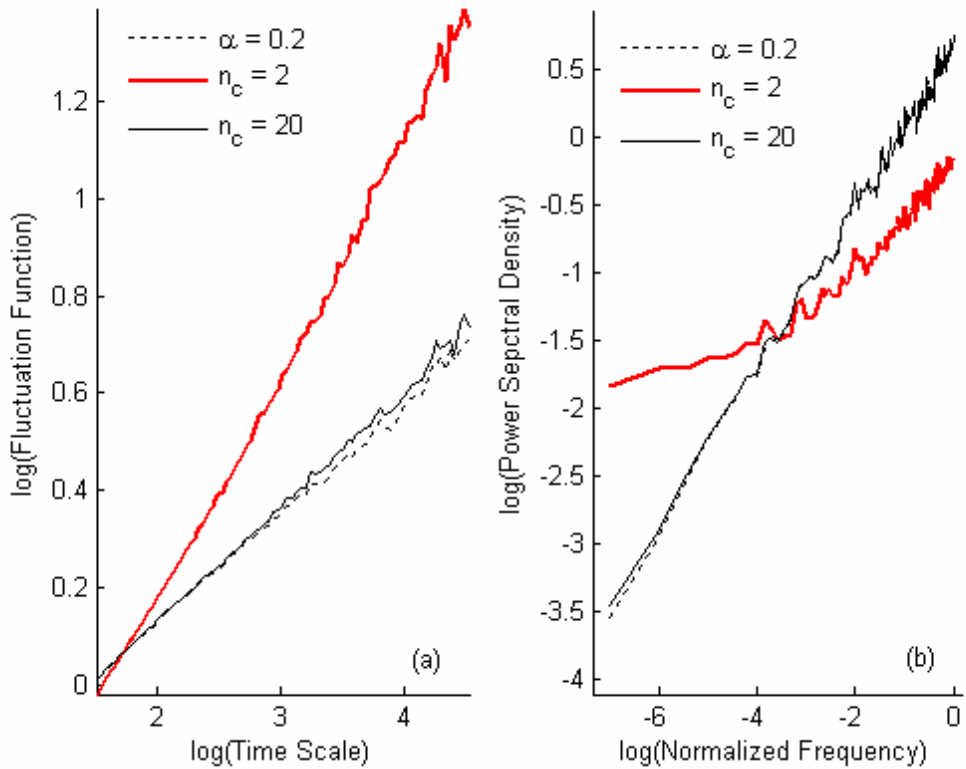

**Figure 4** Log-log plot of the fluctuation function versus time scale of the anti-correlated noise ($\alpha = 0.2$) and that of its coarse-grained versions obtained with $n_c = 2$ (think solid line) and $n_c = 20$ (thin solid line) clusters is shown in (a). The corresponding log-log plot of the power spectrum is shown in (b).

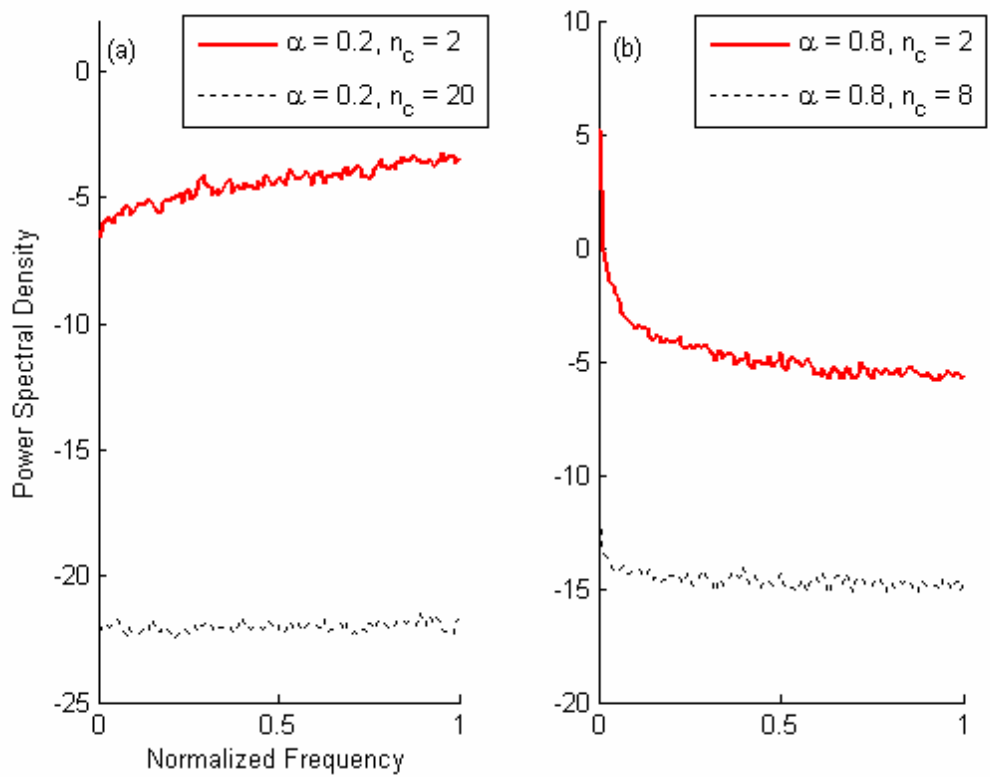

**Figure 5** The power spectrum of the noise $x_n^e$ obtained by clustering the anti-correlated noise ($\alpha = 0.2$) with clusters $n_c = 2$ (solid line) and $n_c = 20$ (dotted line) is shown in (a). Those obtained by clustering the long-range correlated noise ($\alpha = 0.8$) with clusters $n_c = 2$ (solid line) and $n_c = 8$ (dotted line) is shown in (b).

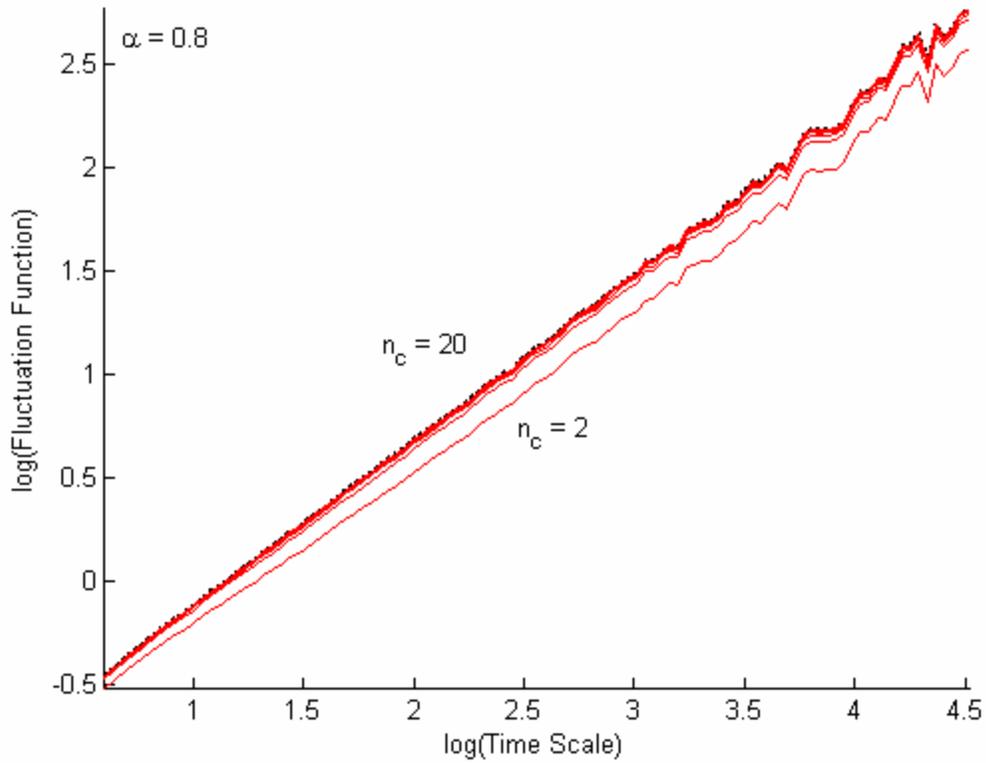

**Figure 6** Log-log plot of the fluctuation function versus time-scale of the long-range correlated noise ($\alpha = 0.8$, dotted line) and its coarse-grained versions generated with ($n_c$ = 2, 4, 6, 8, 10, 12, 14, 16, 18 and 20) clusters (solid lines). The scaling exponent of the coarse-grained realizations exhibit an increasing trend with increasing number of clusters and converge to that of the actual data ($\alpha = 0.8$) for ($n_c$ = 8).

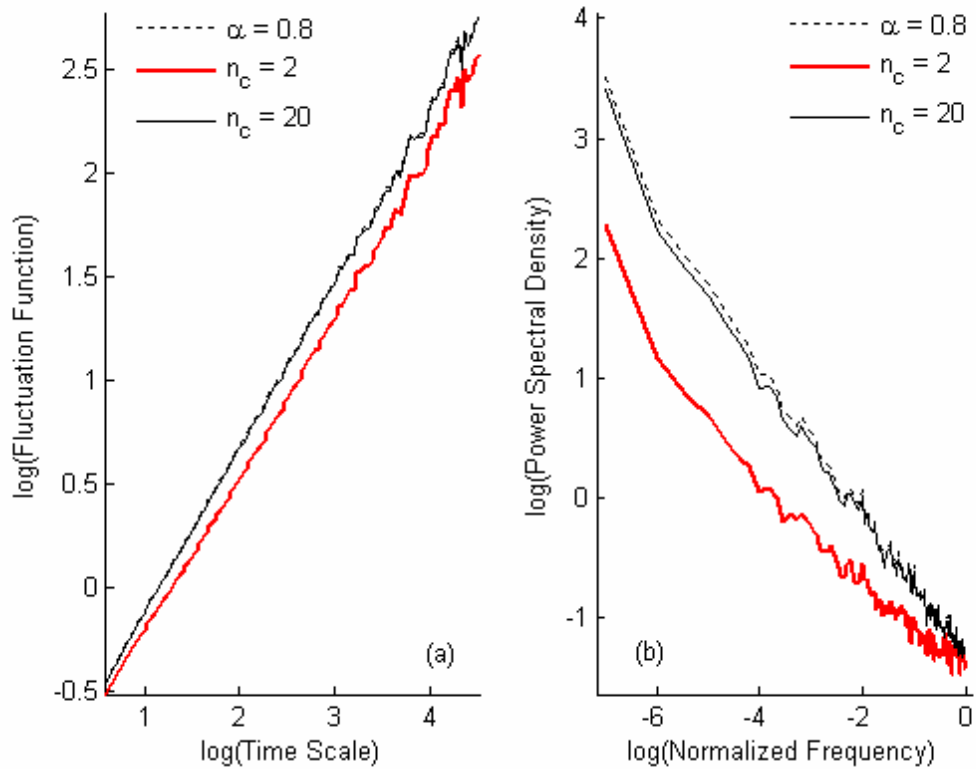

**Figure 7** Log-log plot of the fluctuation function versus time scale of the long-range correlated noise ($\alpha = 0.8$) and that of its coarse-grained versions obtained with $n_c = 2$ (think solid line) and $n_c = 8$ (thin solid line) clusters is shown in (a). The corresponding power spectrum (log-log scale) of these sets is shown in (b).

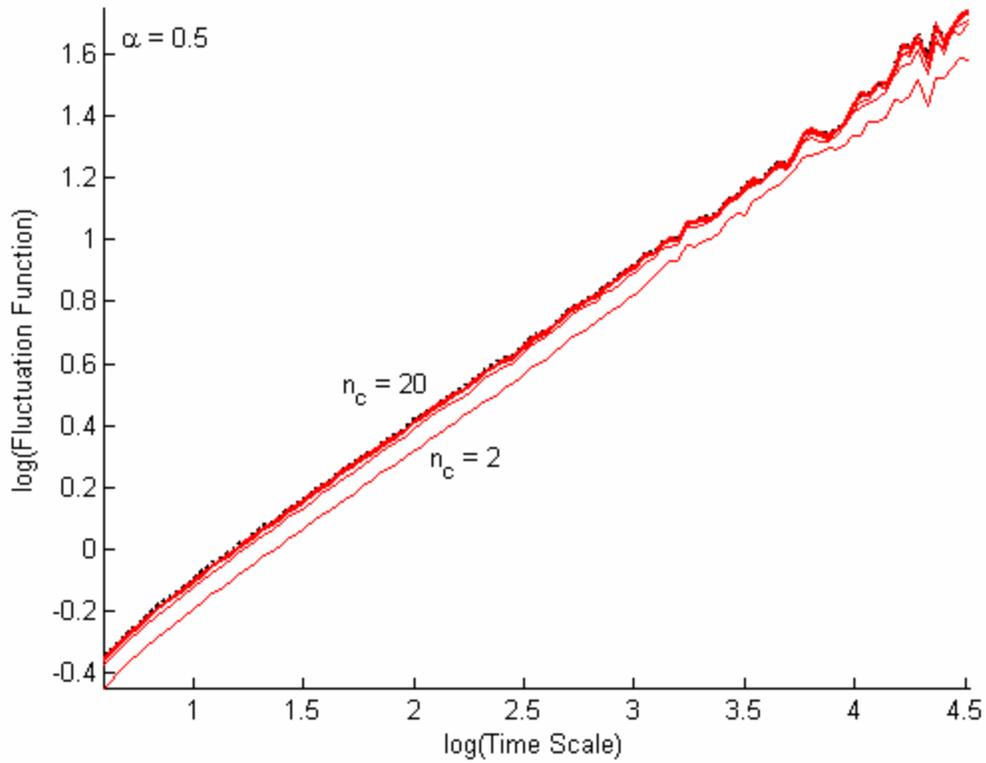

**Figure 8** Log-log plot of the fluctuation function versus time-scale of the uncorrelated noise (α = 0.5, dotted line) and its coarse-grained versions generated with ($n_c$ = 2, 4, 6, 8, 10, 12, 14, 16, 18 and 20) clusters (solid lines). The scaling exponent of the coarse-grained realizations do not exhibit any characteristic trends with increasing number of clusters.

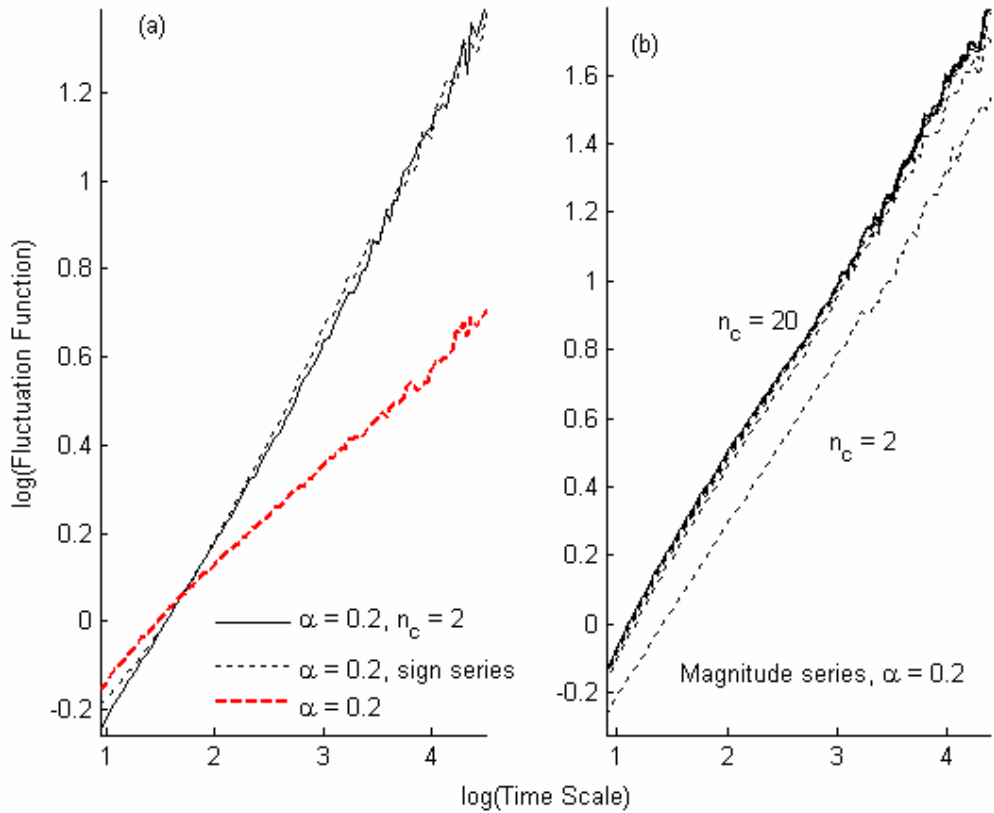

**Figure 9** Log-log plot of the fluctuation function versus time-scale of the anti-correlated noise ($\alpha = 0.2$, thick dashed line), its coarse-grained realization ($n_c = 2$, solid line) along with its sign series (dotted line). Log-log plot of the fluctuation function versus time-scale of the magnitude series of the anti-correlated noise (solid lines) and its coarse-grained realizations ($n_c = 2, 4, 6, 8, 10, 12, 14, 16, 18$ and $20$, dotted lines).

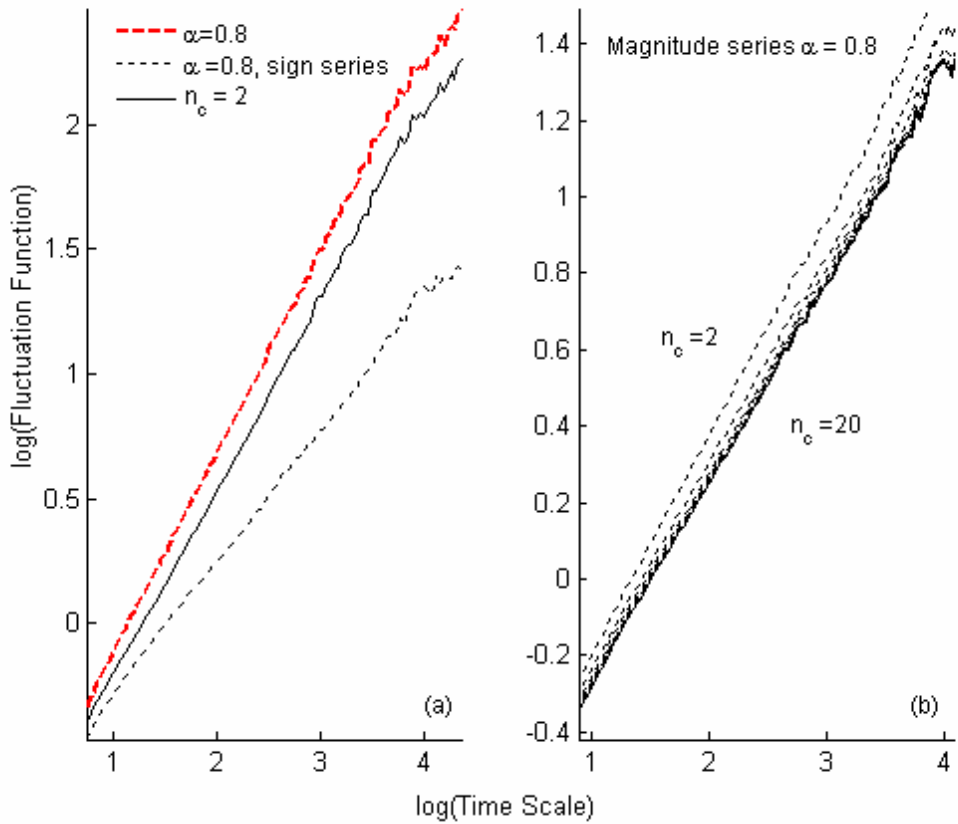

**Figure 10** Log-log plot of the fluctuation function versus time-scale of the long-range correlated noise ($\alpha = 0.8$, thick dashed line), its coarse-grained realization ($n_c = 2$, solid line) along with the sign series (dotted line). Log-log plot of the fluctuation function versus time-scale of the volatility of the long-range correlated noise (solid lines) and volatility of its coarse-grained realizations ($n_c = 2, 4, 6, 8, 10, 12, 14, 16, 18$ and $20$, dotted lines).

**Reference:**

1. Peng C-K. Buldyrev SV. Havlin S. Simons M. Stanley HE. Goldberger AL. Mosaic organization of DNA nucleotides. Phys. Rev. E. 1994; 49: 1685-1689.

2. Proakis JG. and Manolakis DG. *Digital Signal Processing: Principles, Algorithms and Applications* (3rd Edition), Prentice Hall, 1995.

3. MacQueen, J.B. Some Methods for classification and Analysis of Multivariate Observations, Proceedings of 5-th Berkeley Symposium on Mathematical Statistics and Probability. Berkeley, University of California Press. 1967. 1:281-297.

4. Tou, J.T. & Gonzalez R.C. Pattern Recognition Principles (Addison Wesley publishing Co), 1974.

5. Hao B.-L., Elementary symbolic dynamics and chaos in dissipative systems (World Scientific: New Jersey. 1989.

6. Hu K. Ivanov P.C.h. Chen Z. Carpena P. and Eugene Stanley H. Effect of trends on detrended fluctuation analysis. Phys. Rev. E. 2001; 64; 011114 (19 pages).

7. Makhoul, J. Roucos, S. and Gish, H. Vector Quantization in Speech Coding. Proceedings of the IEEE. 1985. 73(11). 1551- 1587.

8. Linde, Y. Buzo, A. and Gray, R.M. An algorithm for vector quantizer design. IEEE Trans. on Comm, 28:84-95, 1980.

9. Meila, M. and Heckerman D. An Experimental Comparison of Several Clustering and Initialization Methods. Microsoft Research Technical Report. 1998. MSR-TR-989-06.

10. Tavare, S. and Giddings, B.W. Some statistical aspects of the primary structure of nucleotide sequences. Mathematical Methods for DNA Sequences. Boca Raton. FL CRC Press, 1989.

11. Peng, C-K. Buldyrev, S.V. Goldberger, A.L. Havlin, S. Sciortino, F. Simon, M. and Stanley, H.E. Long range correlations in Nucleotide Sequences. Nature. 1992. 356:168-170.

12. Ashkenazy, Y. Ivanov, P.Ch. Havlin, S. Peng, C-K. Goldberger, A.L. and Stanley, H.E. Magnitude and Sign Correlations in Heartbeat Fluctuations. Phys.Rev.Lett. 2001. 86(9). 1900-1903.

13. Kantelhardt, J. W. Ashkenazy, Y. Ivanov, P. Ch. Bunde, A. Havlin, S. Penzel, T. Peter, H-J. and Stanley, H.E. Characterization of sleep stages by correlations of magnitude and sign of heartbeat increments. Phys. Rev. E 65. 2002. 051908.

14. Kalisky, T. Ashkenazy, Y. and Havlin, S. Volatility of Linear and Nonlinear Time Series. Physical Review E (in press).